\documentstyle[12pt]{article}
\begin{document}
\noindent
\begin{center}
{\LARGE {\bf  Conformal Anomaly and Large \\Scale Gravitational Coupling\\}}
\vspace{2cm}
${\bf H.~Salehi}$\footnote{e-mail: h-salehi@cc.sbu.ac.ir.}~,
~${\bf Y.~Bisabr}$\footnote{e-mail: y-bisabr@cc.sbu.ac.ir.} \\
\vspace{0.5cm}
{\small {Department of Physics, Sh. Beheshti University, Evin,
Tehran 19839,  Iran.}}\\
\end{center}
\vspace{1cm}
\begin{abstract}
We present a model in which the breackdown of conformal symmetry of a quantum
stress-tensor due to the trace anomaly is related to a cosmological effect in
a gravitational model.  This is done by
characterizing the traceless part of the quantum stress-tensor in terms of the
stress-tensor of
a conformal invariant classical scalar field.  We introduce a conformal frame in
which the anomalous trace is identified with a cosmological constant.  In this
conformal frame we establish the Einstein field equations by connecting the
quantum stress-tensor with the large scale distribution of matter in the
universe.
\end{abstract} \vspace{3cm}
In the absence of a full theory of quantum gravity, one of the
theoretical frameworks in which we may improve our understanding of quantum
processes in a gravitational field is a semiclassical approximation.  In
this framework the matter is described by quantum field theory while
the gravitational field itself is regarded as a classical
object.  The gravitational coupling of a quantum field is then investigated
through the study of the quantum stress-tensor, i.e. the expectation
value of the stress-tensor of the
quantum field taken in some physical state.  However, since
the quantum stress-tensor contains singularities, some
renormalization prescriptions \cite{bd} are used to obtain a meaningful
expression. One
of the most remarkable consequences of these
prescriptions is the so called conformal anomaly.  This means that the trace
of the quantum stress-tensor of a conformal invariant field
obtains a nonzero
expression while the trace of the classical stress-tensor vanishes
identically.  The appearance of a nonvanishing trace may be regarded as
the breackdown of conformal symmetry.  Since the conformal
invariance of a theory reflects its invariance under rescaling
of lengths, one may expect that the anomalous trace of the quantum
stress-tensor may somehow be related to a scale of length.\\
The purpose of this note is to establish this relation by connecting the
trace anomaly with a cosmological constant.  We shall deal with this possibility
by making a distinction between the trace anomaly and the traceless part
of the quantum stress-tensor.  Such a distinction is suggested by the results
of the renormalization theory, although there is an alternative motivation
which can be found in \cite{sal}.  In this way a simple dynamical model is
introduced in which the traceless part is characterized by the
stress-tensor of a conformal invariant scalar field.  In this model, it is
possible to rescale the trace anomaly and convert it into a cosmological
length.  The conformal symmetry
is then broken by this cosmological length and a preferred conformal frame is
determined in which the Einstein field equations with a cosmological
constant are established.\\
Let us begin with the results of renormalization of a quantum stress-tensor
$\Sigma_{\alpha \beta}$ for a quantum scalar field conformally coupled with a
background metric $g_{\alpha\beta}$\cite{wa} \footnote{We
use the units in which $\hbar=c=1$ and follow the sign conventions of Hawking
and Ellis \cite{he}.}

\begin{equation}
\nabla^{\alpha } \Sigma_{\alpha \beta}=0
\label{1}\end{equation}
\begin{equation}
\Sigma_{\alpha }^{\alpha }=-2v_{1}(x)
\label{2}\end{equation}
where
\begin{equation}
v_{1}(x)=\frac{1}{720}\{ \Box R-R_{\alpha\beta}
R^{\alpha\beta}+R_{\alpha\beta\delta\gamma}R^{\alpha\beta\delta\gamma}\}
\label{3}\end{equation}
Here $\nabla_{\alpha }$ denotes a covariant differentiation and
$\Box \equiv g^{\alpha \beta} \nabla_{\alpha } \nabla_{\beta}$; $R_{\alpha \beta \delta \gamma}$ is
the Riemann curvature tensor, $R_{\alpha \beta}$ is the Ricci tensor and $R$ is
the curvature scalar.  The first equation
is a conservation law and the second one
indicates an anomalous trace emerging from the renormalization
process.  The implication
of Eq.(\ref{2}) is that the quantum
stress-tensor $\Sigma_{\alpha \beta}$ may be written in the following
general form
\begin{equation}
\Sigma_{\alpha \beta}=\Sigma_{\alpha \beta}^{(0)}-\frac{1}{2}
g_{\alpha \beta} v_{1}(x)
\label{4}\end{equation}
where $\Sigma_{\alpha \beta}^{(0)}$ is a traceless tensor.  In this
way, $\Sigma_{\alpha \beta}$ is decomposed into two parts: a traceless part
$\Sigma_{\alpha \beta}^{(0)}$ which respects the conformal symmetry
and an anomalous part reflecting the quantum
characteristics.  The traceless condition of $\Sigma_{\alpha \beta}^{(0)}$ is
automatically satisfied if we introduce a conformally invariant C-number
scalar field $\phi$ satisfying
\begin{equation}
(\Box-\frac{1}{6}R)\phi=0
\label{6}\end{equation}
and identify $\Sigma_{\alpha \beta}^{(0)}$ with the
stress-tensor of $\phi$, namely\cite{de}
\begin{equation}
T_{\alpha\beta}[\phi]=(\frac{2}{3}\nabla_{\alpha}\phi\nabla_{\beta}\phi-\frac{1}{6}
g_{\alpha\beta}\nabla_{\gamma}\phi\nabla^{\gamma}\phi)-\frac{1}{3}
(\phi\nabla_{\alpha}\nabla_{\beta}\phi-g_{\alpha\beta} \phi\Box \phi)
+\frac{1}{6}\phi^2 G_{\alpha\beta}
\label{5}\end{equation}
in which $G_{\alpha\beta}$ is the Einstein tensor.
The
relation (\ref{4}) takes then the form
\begin{equation}
\Sigma_{\alpha \beta}=T_{\alpha \beta}-\frac{1}{2}
g_{\alpha \beta} v_{1}(x)
\label{7}\end{equation}
The tracelessness of $T_{\alpha \beta}$ is then ensured by (\ref{6}).\\
We may try to take the relation (\ref{7}) as a general condition
imposed on $\Sigma_{\alpha\beta}$ in the given fixed background geometry
described by the metric tensor $g_{\alpha \beta}$.
However, in this case due to the Eq.(\ref{1}) and the nonvanishing trace
anomaly, $T_{\alpha \beta}$ can not be expressed as a conserved stress-tensor
and the anomalous trace would provide a quantum source
for the traceless tensor $T_{\alpha \beta}$.  Although such a
source could be considered
as a desirable dynamical characteristic in many contexts, but we should note
that its appearance on the whole background metric stands in conflict
with the dynamical equation (\ref{6}) for $\phi$, as a C-number field.   For
this reason we shall follow a different interpretation for the
relation (\ref{7}).\\
We first note that the conformal coupling of the scalar field $\phi$
implies that no distinction can be made among different conformally
related
configurations of $\phi$ and $g_{\alpha\beta}$ in terms of the dynamical
equation (\ref{6}).  Thus, the question which presents
itself is that which conformal frame should be taken as the physical
frame.  We shall choose the conformal frame by the condition
that $T_{\alpha \beta}$ as the stress-tensor of the scalar field, $\phi$,
should actually be conserved.  Following this strategy we consider a conformal
transformation
$$
\bar{g}_{\alpha\beta} =\Omega^{2}(x) g_{\alpha\beta}
$$
\begin{equation}
\bar{\phi}(x) = \Omega^{-1}(x) \phi(x)
\label{9}
\end{equation}
and write Eq.(\ref{7}) in a conformal frame described by the metric
$\bar{g}_{\alpha \beta}$ so that
\begin{equation}
\bar{\Sigma}_{\alpha \beta}=\bar{T}_{\alpha \beta}+\frac{1}{6}
\Lambda \bar{g}_{\alpha \beta} \bar{\phi}^{2}
\label{10}\end{equation}
or, equivalently
\begin{equation}
\bar{G}_{\alpha\beta}+\Lambda \bar{g}_{\alpha\beta}=
6\bar{\phi}^{-2}(\bar{\Sigma}_{\alpha\beta}+\tau_{\alpha\beta}(\bar{\phi}))
\label{12}\end{equation}
where $\Lambda$ denotes a cosmological constant which is taken to be related
to the anomalous trace by
\begin{equation}
-3\bar{\phi}^{-2}\bar{v}_{1}(x)=\Lambda
\label{11}\end{equation}
The coefficient $\bar{\phi}^{-2}$ is introduced to make the dimension of
both sides consistent.  The tensor $\tau_{\alpha\beta}(\bar{\phi})$ is equal
to $\bar{T}_{\alpha \beta}$ without $G_{\alpha \beta}$-term and coincides
up to a sign with the so called modified stress-tensor \cite{pa}.
The relation (\ref{11}) is a constraint on the conformal
factor and singles out a specific conformal frame, which we call the
cosmological frame, in which the anomalous trace
is related to a cosmological constant.
In
the cosmological frame, $\Lambda$ serves to characterize a distinguished
cosmological length
scale which breaks down the conformal symmetry of $\bar{T}_{\alpha \beta}$.
Let us now consider the trace of Eq.(\ref{10})
\begin{equation}
\bar{\Sigma}^{\alpha}_{\alpha } \sim \Lambda  \bar{\phi}^2
\label{13}\end{equation}
Remarkably, this relation permits us to estimate the background average value
of $\bar{\phi}$, if we measure the trace of $\bar{\Sigma}_{\alpha \beta}$
in the cosmological frame in terms of
the large scale distribution of matter\footnote{ This argument seems to be
reasonable since in the cosmological frame the local fluctuations of the
quantum stress-tensor, which are characterized by the trace anomaly, are
replaced by a cosmological constant term.},
$\bar{\Sigma}^{\alpha}_{\alpha } \sim M/R_{0}^3 $ where $M$
and $R_{0}$ are
the mass and the radius of the universe, respectively.  Actually if we take
into account the empirical
fact that the radius of the universe, $R_{0}$, coincides with its Schwarzschild
radius $2GM$, where $G$ is the gravitational constant, the Eq.(\ref{13}) reduces
to
\begin{equation}
\bar{\phi}^{-2} \sim G
\label{14}\end{equation}
provided $\Lambda \sim R_{0}^{-2}$ holds, in agreement with
observations.  Substituting this result into the Eq.(\ref{12}) leads to
\begin{equation}
\bar{G}_{\alpha\beta}+\Lambda \bar{g}_{\alpha\beta} \sim
G \bar{\Sigma}_{\alpha\beta}
\label{14}\end{equation}
which establishes the usual features of the Einstein field equations with a cosmological
constant.  Thus the cosmological frame provides a preferred frame in which
the background average value of $\bar{\phi}$ plays the role of
the gravitational constant.  This result attributes two physical
characteristics to this frame.   Firstly, the
tensor $\bar{T}_{\alpha \beta}$ becomes compatible with
the conservation property of a stress-tensor, and secondly
the large scale gravitational coupling is described by the Einstein field
equations.


\end{document}